\documentclass[a4 paper,12pt]{article}
\usepackage{amsmath}
\usepackage{amsfonts}
\usepackage{amssymb}
\usepackage{graphicx}
\usepackage{array}
\usepackage{multirow}
\usepackage{ulem}
\usepackage{braket}
\usepackage{setspace}
\usepackage[margin=1 in]{geometry}
\usepackage{float}
\usepackage[utf8]{inputenc}
\usepackage{xcolor}
\usepackage{hyperref}
\usepackage{cite}
\newcolumntype{P}[1]{>{\centering\arraybackslash}p{#1}}
\newcolumntype{M}[1]{>{\centering\arraybackslash}m{#1}}
\usepackage{hyperref}

\date{}

\title{}
\begin{document}
\begin{center}
%{\bf\large{BFV quantization and FFBRST of FLPR model} }
{\bf\large{Batalin-Fradkin-Vilkovisky Quantization and Symmetries of FLPR model}}
\vskip 1.5 cm

{\sf{\bf Ansha S. Nair and Saurabh Gupta}}\\
\vskip .1cm
{\it Department of Physics, National Institute of Technology Calicut,\\ Kozhikode - 673 601, Kerala, India}\\

\vskip .15cm
{E-mails: {\tt anshsuk8@gmail.com, saurabh@nitc.ac.in}}
\end{center}
\vskip 1cm
\noindent
\textbf{Abstract:} 
We quantize the Friedberg-Lee-Pang-Ren (FLPR) model within the framework of Batalin-Fradkin-Vilkovisky (BFV) formalism. We construct the nilpotent Becchi-Rouet-Stora-Tyutin (BRST) charges using constraints and the fermionic gauge-fixing function by means of admissible gauge conditions. We also derive the BRST invariant effective action (and corresponding symmetries) of the model in both polar and Cartesian coordinates. We demonstrate that the physical states of the system are annihilated by the first-class constraints which is consistent with the Dirac formalism. Moreover, we establish the finite field-dependent BRST (FFBRST) symmetries of the FLPR model. We exhibit the interlink between the BFV-BRST gauge-fixed action and the classical gauge invariant action using FFBRST formulation. 

\vskip 1.5cm
\noindent
\textbf{PACS:} 11.15.-q, 11.10.Ef, 11.30.-j
\vskip 1cm
\noindent
\textbf{Keywords:} FLPR model; BFV formalism; BRST and FFBRST symmetries 

\clearpage

\section{Introduction}
Constrained theories are the cornerstone of modern theoretical   physics, as they include gauge theories, gravity theories, super-symmetric theories etc. However, the conventional methods of quantization are not directly applicable to such theories. To this end, Dirac (and independently Anderson $\&$ Bergmann) developed a quantization technique -  popularly known as the Dirac formalism  - to deal with such constrained systems \cite{1,And}.  In Dirac formalism, constraints are categorized into primary, secondary, tertiary etc. and further classified into first-class and second-class. Later, Faddeev and Jackiw came up with a formalism where all the constraints are treated on equal footing without any classification \cite{2}.  The  Faddeev-Jackiw (FJ) formalism is a geometrically motivated approach and based on the symplectic structure of the phase space \cite{Bar}. This formalism has gained a lot of interest recently and being used widely (see e.g. \cite{FJ1,FJ2,FJ3, FJ4} and references therein).

On the other hand, to quantize such constrained systems the BRST-BFV formalism \cite{3,4,BFV1,BFV2} is one of the most sought after methods. It provides a systematic Hamiltonian framework to quantize a theory endowed with first-class constraints and also preserves the BRST symmetries of the system. The BFV formalism has also been generalized to include second-class theories by employing a conversion technique called the BFT formalism \cite{BFT1}. These methods have been explored in a variety of classical and field theoretical models such as Chiral Schwinger model, Proca theory, QED and $(2+1)$-dimensional Ho{\v{r}}ava theory are the few names \cite{BFV3,BFV4,BFV5,BFV6}.

The generalization of the BRST symmetries by making the anti-commuting infinitesimal transformation parameter finite and field dependent is known as the finite field dependent BRST (FFBRST) transformations, introduced in \cite{FF1}. The effective action is invariant under such transformations; however the field dependent nature of transformation parameter disrupts the symmetry of the path integral measure. The non-trivial change in the Jacobian of the path integral measure leads to the emergence of interesting and novel results. Consequently, with a judicious choice of the finite field dependent parameter, FFBRST transformations relate the generating functional corresponding to two different actions \cite{FF2,FF3}. This generalized BRST (FFBRST) symmetry transformation has found numerous applications in systems such as Yang-Mills theory, Bagger-Lambert Gustavsson (BLG) theory in $N=1$ superspace, ABJM theory in $N=1$ superspace, spin-s conformal field theories, fourth-order Pais-Uhlenbeck oscillator (PUO), Christ-Lee model \cite{FF4,FF5,FF6,FF7,FF8,FF9,FF10,FF11,FF12,FFNPB} etc. Nevertheless, some basic and intriguing models, such as the FLPR model, have yet to be explored using FFBRST transformations.  

The FLPR model is a soluble gauge model that describes the motion of a single non-relativistic particle of unit mass governed by a central potential \cite{FLPR1}. It is a $(0+1)$-dimensional quantum mechanical model with the characteristics of Gribov ambiguity and inspired by the dimensional reduction of Yang-Mills theory \cite{FLPR5}. In order to avoid the gauge-fixing procedure this model is being explored using a physical projector operator \cite{FLPR4}. On the other hand, the BRST invariant path integral is also formulated by summing over all Gribov-type copies \cite{FLPR2}. The FLPR model is being studied using an admissible gauge-fixing condition within the framework of FJ formalism \cite{FLPR3}. Whereas the existence of some {\it novel} off-shell nilpotent and absolutely anti-commuting symmetries of the FLPR model is established in \cite{FLPR5}.  This is an interesting model which is also being investigated by means of the supervariable approach (cf. \cite{FLPR6,FLPR8,FLPR9} for details). Recently, a consistent quantization of FLPR model by avoiding the Gribov copies with corresponding BRST related symmetries is performed in \cite{FLPR10}.

The prime motive of our present endeavor is to quantize the FLPR model using an admissible gauge-fixing condition, in both polar and Cartesian coordinates, by employing BFV formalism. We intend to construct the nilpotent and conserved BRST charge and subsequently derive the BRST invariant gauge-fixed effective action and corresponding symmetries. Additionally, we aim to explore physicality criteria with respect to the conserved BRST charges and compare vis-$\grave{a}$-vis Dirac-quantization conditions. Finally, we wish to establish the FFBRST symmetries and thereby connect the gauge-fixed effective action and classical gauge invariant action of the FLPR model.

The content of the present manuscript is organized as follows. In the second section, we briefly review the BFV formalism. We provide the constraint structure of the FLPR model, in polar coordinates, using Dirac's prescription and carried out BFV quantization in the third section.  Our subsequent section deals with the constraint analysis and BFV quantization of the FLPR model in Cartesian coordinates. The conserved and nilpotent BRST charges and the physicality criteria are described in Section 5. Our section 6 deals with the FFBRST formalism of the FLPR model in polar coordinates. Finally, we provide some concluding remarks. 

\section{BFV Formalism}
BFV formalism provides a Hamiltonian framework for quantization of gauge theories by enlarging the phase space with ghost variables and preserves the BRST symmetry.

Consider a Hamiltonian $H_{c}(q_{i},p_{i})$ defined in a $2n$-dimensional phase space of canonical variables $q_{i},\,p_{i}$ $(i=1,2,...,n)$, where $q_{i}$ are generalized coordinates and $p_{i}$ are generalized momenta. We assume that the constraints $\Omega_{a}(q_{i},p_{i})$ ($a=1,2,...,m$) are irreducible and adhere to the following constraint algebra\cite{BFV3}
\begin{equation}
    \big[\Omega_{a},\,\Omega_{b}\big]=i\,\Omega_{c}U_{ab}^{c},\qquad \big[H_{c},\,\Omega_{b}\big]=i\,\Omega_{a}V_{b}^{a},
\end{equation}
where $U_{ab}^{c}$ and $V_{b}^{a}$ are structure coefficients. We introduce additional conditions, known as gauge-fixing conditions $\Psi^{a}(q^{i},p_{i})\approx 0$, for the purpose of isolating the physical variables in the theory. The condition $\Psi^{a}$ defines a surface that cuts through every gauge-orbit once and only once. It chooses a single representative from each gauge orbit and eliminates redundant gauge degrees of freedom. Now, we  obtain $2(n-m)$-dimensional phase space defined by the physical canonical variables, say $q^{*}$ and $p^{*}$. The theory of described dynamical system depends only on the canonical variables $q^{*}$ and $p^{*}$. Thus, the generating functional is given by \cite{BFV3}
\begin{equation}
    {\cal{Z}}=\int \big[dq^{*}dp^{*}\big]\exp\Bigg[i\int dx \Big(p^{*}\dot{q}^{*}-H_{phys}(q^{*},p^{*})\Big)\Bigg],
\end{equation}
here, $H_{phys}$ represents the Hamiltonian defined in terms of the physical variables in the theory. The action containing constraints $\Omega_{a}\approx 0$ and the gauge-fixing condition $\Psi^{a}(q^{i},p_{i})\approx 0$ can be written as
\begin{equation}
    S=\int dt\,\Big(p_{i}\dot{q}^{i}-H_{c}-\lambda^{a}\Omega_{a}+\pi_{a}\Psi^{a}\Big),
\end{equation}
where $\lambda^{a}$ and $\pi_{a}$ are Lagrange multiplier canonically conjugate to each other. The gauge-fixing condition generally takes the following form
\begin{equation}
    \Psi^{a}=\dot{\lambda}^{a}+\chi^{a}(q^{i},p_{i},\lambda^{a}),
\end{equation}
where $\chi^{a}$ are arbitrary functions. Here, we observe that $\lambda^{a}$ become dynamically evolving and $\pi_{a}$ act as their conjugate momenta. This leads to the canonical formalism in an extended phase space.

In order to make use of the BFV formalism, we extend the phase space by introducing two sets of canonical ghost and anti-ghost variables $\big({\cal{C}}^{a},{\cal\bar{P}}_{a}\big)$ and $\big({\cal{P}}^{a},{\cal\bar{C}}_{a}\big)$ along with the auxiliary variables $ \big({N}^{a},{B}_{a}\big)$ for each constraint in the theory, such that they satisfy the following super-Poisson bracket algebra \cite{BFV1,BFV2}
\begin{equation}\label{sp}
\begin{gathered}
    \{{\cal{C}}^{a},{\cal\bar{P}}_{b}\}=i\,\delta_{b}^{a}, \quad
    \{{\cal{P}}^{a},{\cal\bar{C}}_{b}\}=i\,\delta_{b}^{a},\quad
  \big[{N}^{a},{B}_{b}\big]=i\,\delta_{b}^{a}.   
\end{gathered}
\end{equation}
The super-Poisson bracket between any two variables $A$ and $B$ is defined as
\begin{equation}\label{n3}
    \{A,B\}=i\,\Bigg(\frac{\partial_{r}A}{\partial q}\frac{\partial_{l}B}{\partial p}-(-1)^{\eta_{A}\eta_{B}}\frac{\partial_{r}B}{\partial q}\frac{\partial_{l}A}{\partial p}\Bigg),
\end{equation}
where $\eta_{A}$($\eta_{B}$) equals $0$ or $1$ depending on whether $A(B)$ is a boson or fermion, respectively and the subscript $r$ and $l$ denotes the right and left derivatives, respectively \cite{BFV3}. It is important to note that $\big[A,B\big]=i\,\{A,B\}$. A non-trivial property of Eq. \eqref{n3} is that if $A$ is a fermion, then \cite{BFV2}
\begin{equation}
    \{A,A\}\neq 0, \qquad \big\{\{A,\,A\},\,A\big\}=0.
\end{equation}
In the extended phase space, the nilpotent BRST charge $(Q)$ takes the following form:
\begin{equation}
    Q={\cal{C}}^{a}\Omega_{a}\,+\,{\cal{P}}^{a}B_{a}.
\end{equation}
The gauge-fixing function $(\Phi)$ can be constructed as 
\begin{equation}
    \Phi={\cal\bar{C}}_{a}\,\chi^{a}+{\cal\bar{P}}_{a}\,N^{a},
\end{equation}
where $\chi^{a}$ are the gauge-fixing conditions of the theory. The minimal Hamiltonian $(H_{m})$ is defined by 
\begin{equation}
   H_{m}=H_{c}+ \bar{{\cal{P}}_{a}}V^{a}_{b}{\cal{C}}^{b}. 
\end{equation}
At this juncture, it is worth mentioning that the BRST charge $(Q)$ is conserved and fermionic in nature. As it is straightforward to verify that \cite{BFV4}
\begin{equation}\label{11}
  \big[Q,\;H_{m}\big]=0, \qquad
    Q^{2}=\{Q,\;Q\}=0.  
\end{equation}
In addition, the fermionic gauge-fixing function $(\Phi)$ and the BRST charge $(Q)$ satisfy the following algebra
\begin{equation}\label{e12}
     \{\{\Phi,\;Q\}, \;Q\}=0.
\end{equation}
The quantum theory is defined by the generating functional in the extended phase space as \cite{Henn}
\begin{equation}
    {\cal{Z}}=\int \big[{\cal{D}}\phi\big]e^{iS_{eff}},
\end{equation}
where $\big[{\cal{D}}\phi\big]$=\,$[dq^{i}\,dp_{i}]\,[d{\cal{C}}^{a}\,d{\cal\bar{P}}_{a}]\,[d{\cal{P}}^{a}\,d{\cal\bar{C}}_{a}]\,[d{N}^{a}\,d{B}_{a}]$ and ``effective action'' $(S_{eff})$ is given by
\begin{eqnarray}
     S_{eff}=\int dt\,\Big[ p_{i}\,\dot{q}^{i}+B_{a}\,\dot{N^{a}}+{\cal\bar{P}}_{a}\,{\cal{\dot{C}}}^{a}+{\cal\bar{C}}_{a}\,{\cal{\dot{P}}}^{a}-H_{total}\Big],
\end{eqnarray}
where $H_{total}=H_{m}-i\,\{\Phi,\;Q\}$.

\section{FLPR model in polar coordinates}\label{sec2}
We begin with the Lagrangian that describes the dynamics of the FLPR model in polar coordinates as \cite{FLPR1}
\begin{equation}\label{pol}
       {L}=\frac{1}{2}\dot{r}^{2}+\frac{1}{2}r^{2}(\dot{\theta}-g\zeta)^{2}+\frac{1}{2}(\dot{z}-\zeta)^{2}-V(r),
\end{equation}
where $\dot{r}$, $\dot{\theta}$ and $\dot{z}$ are generalized velocities corresponding to the generalized coordinates $r$, $\theta$ and $z$, respectively. Here, $\zeta$ represents a gauge variable and $g$ is  coupling constant.

\subsection{Constraint structure}
The canonical momenta corresponding to the variables $r$, $\theta$, $z$ and $\zeta$ are given, respectively, as
\begin{equation}\label{2}
    p_{r}=\dot{r},\quad
    p_{\theta}=r^{2}(\dot{\theta}-g\zeta),\quad
    p_{z}=\dot{z}-\zeta,\quad p_{\zeta}=0.
\end{equation} 
We identify $\Omega_{1}\equiv p_{\zeta}\approx 0$ as the primary constraint in the theory \textit{\`{a} la} Dirac's prescription for classification of constraints \cite{1}. 
The canonical Hamiltonian $(H_{c})$  can be obtained as follows
\begin{equation}
   {H}_{c}= \frac{p_{r}^{2}}{2}+\frac{p_{\theta}^{2}}{2r^{2}}+\frac{p_{z}^{2}}{2}+\zeta(gp_\theta+p_{z})+V(r).
\end{equation}
The consistency condition on primary constraint $\Omega_{1}$ leads to a secondary constraint $(\Omega_{2})$
\begin{equation}
    \Omega_{2}\equiv gp_{\theta}+p_{z}\approx 0.
\end{equation}
There are no further constraints in the theory. Thus, the theory possess only two constraints, which are first-class in nature.
This indicates that the underline theory exhibit a gauge symmetry. The corresponding gauge transformations are listed below: (cf. \cite{FLPR3} for details)
\begin{eqnarray}
%\begin{split}
    &\Tilde{\delta} r=0,\qquad \Tilde{\delta} P_{r}=0,\qquad
    \Tilde{\delta} \theta=-g\lambda(t),\nonumber\\
    &\Tilde{\delta} P_{\theta}=0,\qquad
    \Tilde{\delta} z=-\lambda(t),\qquad \Tilde{\delta} P_{z}=0, \qquad \Tilde{\delta}\zeta=-\dot{\lambda}(t).
    %\end{split}
\end{eqnarray}
It is straightforward to verify that the Lagrangian in Eq.~\eqref{pol} remains invariant under the above set of gauge transformations.

\subsection{BFV quantization}\label{sec2.1} We introduce two sets of canonical ghost and anti-ghost variables $\big({\cal{C}}^{a},{\cal\bar{P}}_{a}\big)$ and $\big({\cal{P}}^{a},{\cal\bar{C}}_{a}\big)$ along with the auxiliary variables $ \big({N}^{a},{B}_{a}\big)$, where $a=1,2.$
According to the BFV formalism, the nilpotent BRST charge $(Q)$ is constructed using constraints as follows 
\begin{equation}
   Q={\cal{C}}^{1}\,p_{\zeta}+{\cal{C}}^{2}\,(gp_{\theta}+p_{z})+{\cal{P}}^{1}\,B_{1}+{\cal{P}}^{2}\,B_{2},
   \end{equation}
and the gauge-fixing function $(\Phi)$ is defined as
   \begin{equation}
    \Phi={\cal\bar{C}}_{1}\,\chi^{1}+{\cal\bar{C}}_{2}\,\chi^{2}+{\cal\bar{P}}_{1}\,N^{1}+{\cal\bar{P}}_{2}\,N^{2}.
\end{equation}
Here, we have chosen the admissible gauge-fixing conditions $\chi^{1}=\zeta$ and $\chi^{2}=z-\frac{1}{2}\,B_{2}.$ It is worth mentioning that the other gauge-fixing conditions are plagued by Gribov ambiguities (cf. \cite{FLPR1} for details).

The minimal Hamiltonian $(H_{m})$ takes the following form
\begin{eqnarray}
    H_{m}=\frac{p_{r}^{2}}{2}+\frac{p_{\theta}^{2}}{2r^{2}}+\frac{p_{z}^{2}}{2}+\zeta\,(gp_\theta+p_{z})+V(r)+{\cal\bar{P}}_{2}\,{\cal{C}}^{1}.
\end{eqnarray}
Now, the generating functional in the extended phase space can be written as
\begin{eqnarray}\label{12}
    {\cal{Z}}=\int {\cal{D}}\phi\,\exp\Big(i\,\int dt\,\Big[ p_{r}\,\dot{r}+p_{\theta}\,\dot{\theta}+p_{z}\,\dot{z}+p_{\zeta}\dot{\zeta}+B_{a}\,\dot{N^{a}}+{\cal\bar{P}}_{a}\,{\cal{\dot{C}}}^{a}+{\cal\bar{C}}_{a}\,{\cal{\dot{P}}}^{a}-H_{total}\Big],
\end{eqnarray}
where $ {\cal{D}}\phi={\cal{D}}\big(r, p_{r},\theta, p_{\theta}, z, p_{z}, \zeta, p_{\zeta}, N_{a}, B_{a}, {\cal{C}}^{a}, \bar{{\cal{P}}}_{a}, {\cal{P}}^{a}, \bar{{\cal{C}}}_{a}\big)$
and $H_{total}=H_{m}-i\,\{\Phi,\;Q\}$. We replace $\chi^{1}$ with $\chi^{1}+\dot{N}^{1}$ and subsequently omit the terms
$\int dt\, (B_{1}\,\dot{N^{1}}\,+\,{\cal\bar{C}}_{1}\,{\cal{\dot{P}}}^{1}\,)\,= \,\,\{Q, \int dt\,{\cal\bar{C}}_{1}\,\dot{N}^{1}\}$ in the above Eq. \eqref{12}. Thus, the effective action takes the following form:
\begin{eqnarray}
    S_{eff}=\int dt\,\Big[ p_{r}\,\dot{r}+p_{\theta}\,\dot{\theta}+p_{z}\,\dot{z}+p_{\zeta}\dot{\zeta}+B_{2}\,\dot{N^{2}}+{\cal\bar{P}}_{1}\,{\cal{\dot{C}}}^{1}+{\cal\bar{P}}_{2}\,{\cal{\dot{C}}}^{2}+{\cal\bar{C}}_{2}\,{\cal{\dot{P}}}^{2}-H_{total}\Big].
\end{eqnarray}
In order to obtain the covariant effective action, we first eliminate the variables $B_{1}$, $N^{1}$, ${\cal\bar{C}}_{1}$, ${\cal{P}}^{1}$,  ${\cal\bar{P}}_{1}$, ${\cal{C}}^{1}$, $\zeta$ and $p_{\zeta}$ by using Gaussian integration. Consequently, the explicit form of the effective action turns out to be 
\begin{eqnarray}\label{13}
    S_{eff}&=&\int dt\, \Big[p_{r}\,\dot{r}+p_{\theta}\,\dot{\theta}+p_{z}\,\dot{z}+B\,\dot{N}+{\cal\bar{P}}\,{\cal{\dot{C}}}+{\cal\bar{C}}\,{\cal{\dot{P}}}-\frac{1}{2}\,p_{r}^{2}-\frac{1}{2r^{2}}\,p_{\theta}^{2}-\frac{1}{2}\,p_{z}^{2}\nonumber\\&-&V(r)-(gp_{\theta}+p_{z})\,N-{\cal\bar{C}}\,{\cal{C}}-B\Big(z-\frac{1}{2}\,B \Big)-{\cal\bar{P}}\,{\cal{P}}\Big],
\end{eqnarray}
with $N^{2}\equiv N$, $B_{2}\equiv B$, ${\cal\bar{C}}_{2}\equiv{\cal\bar{C}}$, ${\cal{C}}^{2}\equiv{\cal{C}}$,  ${\cal\bar{P}}_{2}\equiv{\cal\bar{P}}$ and   ${\cal{P}}^{2}\equiv{\cal{P}}$. Second, the variation with respect to $p_{r}$, $p_{\theta}$, $p_{z}$, ${\cal{P}}$ and ${\cal\bar{P}}$ yields the following relations
\begin{eqnarray}\label{14}
   &&p_{r}=\dot{r}, \qquad
   p_{\theta}=r^2\,(\dot{\theta}-gN),\qquad
   p_{z}=\dot{z}-N, \qquad {\cal\bar{P}}=-{\dot{\bar{\cal{C}}}},\qquad
   {\cal{P}}=\dot{\cal{C}}.
\end{eqnarray}
Finally, substituting Eqs. \eqref{14} into the action \eqref{13} and identifying $N$ with $\zeta$ (cf. \eqref{2}), we obtain the covariant effective action as
\begin{eqnarray}\label{15}
    S_{eff}&=&\int dt\, \Big[\frac{1}{2}\,\dot{r}^{2}+\frac{1}{2}\,r^{2}\big(\dot{\theta}-g\zeta \big)^{2}+\frac{1}{2}\big(\dot{z}-\zeta\big)^{2}-V(r)\nonumber\\&+&B\Big(\dot{\zeta}-z+\frac{1}{2}\,B\Big)-{\dot{\bar{\cal{C}}}}\,{\dot{\cal{C}}}-{\cal{\bar{C}}}\,{\cal{C}}\Big].
\end{eqnarray}
This $(S_{eff})$ is invariant under the following off-shell nilpotent BRST symmetry transformations $(s_b)$
\begin{eqnarray}\label{BRST1}
    s_{b} r=0,\quad
    s_{b} \theta=g{\cal{C}},\quad s_{b} z={\cal{C}},
      \quad s_{b}\zeta=\dot{\cal{C}},\quad
   s_{b}{\cal{C}}=0,\quad s_{b}{\cal\bar{C}}=-B,\quad s_{b} B=0,
\end{eqnarray}
as $s_{b} L_{eff}=\frac{d}{dt}(B\dot{\cal{C}})$.

At this juncture, it should be noted that if we take variation over $\cal{C}$ and $\cal\bar{C}$ instead of $\cal{P}$ and $\cal\bar{P}$ in Eq. \eqref{13}, we obtain the following effective action
\begin{eqnarray}\label{17}
    S^\prime_{eff}&=&\int dt\, \Big[\frac{1}{2}\,\dot{r}^{2}+\frac{1}{2}\,r^{2}\big(\dot{\theta}-g\zeta \big)^{2}+\frac{1}{2}\big(\dot{z}-\zeta\big)^{2}-V(r)\nonumber\\&+&B\Big(\dot{\zeta}-z+\frac{1}{2}\,B\Big)-{\dot{\bar{\cal{P}}}}\,{\dot{\cal{P}}}-{\cal\bar{P}}\,{\cal{P}}\Big].
\end{eqnarray}
This form of effective action ($S^\prime_{eff}$) can be obtained by the variable transformation
$\cal{C}\longrightarrow\cal{P}$ and
$\cal\bar{C}\longrightarrow\cal\bar{P}$ in Eq. \eqref{15} and vice-versa.

\section{FLPR model in Cartesian coordinates} \label{sec3}
The Lagrangian that describes the FLPR model in Cartesian coordinates is given by \cite{FLPR1, FC}
\begin{equation}\label{car}
    \Tilde{L}=\frac{1}{2}\Big[(\dot{x}+gy\zeta)^{2}+(\dot{y}-gx\zeta)^{2}+(\dot{z}-\zeta)^{2}\Big]-U(x^{2}+y^{2}),
\end{equation}
where $\dot{x}$, $\dot{y}$ and $\dot{z}$ represent the generalized velocities, $\zeta$ denotes the gauge variable and $g$ is a coupling constant. The formulation of the FLPR model in polar and Cartesian coordinates are {\it physically} equivalent. These coordinates are related through the standard transformations, as: $x=r\cos{\theta}\,\,\text{and}\,\, y=r\sin{\theta}$ (cf. \cite{FLPR4} for discussion on this aspect).

\subsection{Constraint structure}
The canonical momenta corresponding to the coordinates $x$, $y$, $z$ and $\zeta$ are obtained as follows:
\begin{equation}\label{21}
    p_{x}=\dot{x}+gy\zeta, \quad p_{y}=\dot{y}-gx\zeta, \quad p_{z}=\dot{z}-\zeta, \quad p_{\zeta}=0.
\end{equation}
We identify $\Tilde{\Omega}_{1}\equiv p_{\zeta}\approx 0$ as a primary constraint in the theory. The canonical Hamiltonian $({\Tilde{H}_{c}})$ of the system is given as
\begin{equation}
    \Tilde{H}_{c}=\frac{1}{2}\Big(p_{x}^{2}+p_{y}^{2}+p_{z}^{2}\Big)+\zeta\Big(g(xp_{y}-yp_{x})+p_{z}\Big)+U(x^{2}+y^{2}).
\end{equation}
The time evolution of the primary constraint yields a secondary constraint $(\Tilde{\Omega}_{2})$ as follows:
\begin{equation}
    \Tilde{\Omega}_{2}=g\,(xp_{y}-yp_{x})+p_{z}.
\end{equation}
There are only two constraints $(\Tilde{\Omega}_{1}$ and $\Tilde{\Omega}_{2})$ in the theory and these are first-class in nature.
Hence, the Lagrangian in Eq.~\eqref{car} respects the following set of gauge transformations  (cf. \cite{FLPR3} for details):
\begin{eqnarray}\label{gtc}
%\begin{split}
    &\delta x=gy\lambda(t),\qquad \delta P_{x}=gP_{y}\lambda(t),\qquad
    \delta y=-gx\lambda(t),\nonumber\\
    &\delta P_{y}=-gP_{x}\lambda(t),\qquad
    \delta z=-\lambda(t),\qquad \delta P_{z}=0, \qquad \delta\zeta=-{\dot{\lambda}(t)}.
    %\end{split}
\end{eqnarray}
where $\lambda$ is a time infinitesimal gauge parameter. It is straightforward to verify the invariance of Lagrangian ($\tilde L$) under the above gauge transformations.

\subsection{BFV quantization}\label{sec3.1}
As per the BFV formulation, the nilpotent BRST charge ($\tilde{Q}$) and the fermionic gauge-fixing function ($\tilde{\Phi}$) are defined using constraints and gauge-fixing conditions, respectively, as follows: 
\begin{eqnarray}
   \Tilde{Q}&=&{\cal{C}}^{1}\,p_{\zeta}+{\cal{C}}^{2}\,\big(g\,(xp_{y}-yp_{x})+p_{z}\big)+{\cal{P}}^{1}\,B_{1}+{\cal{P}}^{2}\,B_{2},\\
\Tilde{\Phi}&=&{\cal\bar{C}}_{1}\,\Tilde{\chi}^{1}+{\cal\bar{C}}_{2}\,\Tilde{\chi}^{2}+{\cal\bar{P}}_{1}\,N^{1}+{\cal\bar{P}}_{2}\,N^{2},
\end{eqnarray}
where $\Tilde{\chi}^{1}=\zeta$ and $\Tilde{\chi}^{2}=z-\frac{1}{2}B_{2}$ are the admissible gauge-fixing conditions. The minimal Hamiltonian $(\Tilde{H}_{m})$ takes the following form
\begin{eqnarray}
    \Tilde{H}_{m}=\frac{1}{2}\Big(p_{x}^{2}+p_{y}^{2}+p_{z}^{2}\Big)+\zeta\Big(g(xp_{y}-yp_{x})+p_{z}\Big)+U(x^{2}+y^{2})+{\cal\bar{P}}_{2}\,{\cal{C}}^{1}.
\end{eqnarray}
Moreover, $\Tilde{Q}$, $\Tilde{\Phi}$ and $\Tilde{H}_{m}$ satisfy the algebra as listed in Eqs. \eqref{11} and \eqref{e12}.

Hence, the generating functional in the extended phase space can be given as
\begin{equation}
    \Tilde{\cal{Z}}=\int{\cal{D}}\Tilde{\phi}\,\exp\big(i\,\Tilde{S}_{eff}
    \big),
    \end{equation}
    where ${\cal{D}}\Tilde{\phi}={\cal{D}}\big(x, p_{x},y, p_{y}, z, p_{z}, \zeta, p_{\zeta}, N_{a}, B_{a}, {\cal{C}}^{a}, \bar{{\cal{P}}}_{a}, {\cal{P}}^{a}, \bar{{\cal{C}}}_{a}\big)$ and
\begin{eqnarray}
    \Tilde{S}_{eff}=\int dt \Big[p_{x}\dot{x}+p_{y}\dot{y}+p_{z}\dot{z}+p_{\zeta}\dot{\zeta}+B_{a}\dot{N^{a}}+{\cal\bar{P}}_{a}{\cal{\dot{C}}}^{a}+{\cal\bar{C}}_{a}{\cal{\dot{P}}}^{a}-\Tilde{H}_{total}\Big].
\end{eqnarray}
 Now, we suppress the terms $\int dt\, (B_{1}\,\dot{N^{1}}$+\,${\cal\bar{C}}_{1}\,{\cal{\dot{P}}}^{1}\,)\,$=$\,\,\{\,\Tilde{Q},\int dt\,\;{\cal\bar{C}}_{1}\,\dot{N}^{1}\}$ in the Legendre transformations by replacing $\Tilde{\chi}^{1}$ with $\Tilde{\chi}^{1}+\dot{N}^{1}$. Later, in order to obtain the effective action, first we perform the path integration over the variables $B_{1}$, $N^{1}$, ${\cal\bar{C}}_{1}$, ${\cal{P}}^{1}$,  ${\cal\bar{P}}_{1}$, ${\cal{C}}^{1}$, $\zeta$ and $p_{\zeta}$ using Gaussian integration. Thus, we obtain
\begin{eqnarray}\label{27}
    \Tilde{S}_{eff}&=&\int dt \Big[p_{x}\dot{x}+p_{y}\dot{y}+p_{z}\dot{z}+B\dot{N}+{\cal\bar{P}}{\cal{\dot{C}}}+{\cal\bar{C}}{\cal{\dot{P}}}- \frac{p_{x}^{2}}{2}-\frac{p_{y}^{2}}{2} -\frac{p_{z}^{2}}{2} -U(x^{2}+y^{2})\nonumber\\&-&\big(g(xp_{y}-yp_{x})+p_{z}\big)N-B(z-\frac{1}{2}B)-{\cal\bar{C}}{\cal{C}}-{\cal\bar{P}}{\cal{P}}\Big],
\end{eqnarray}
here, we have redefined $N^{2}\equiv N$, $B_{2}\equiv B$, ${\cal\bar{C}}_{2}\equiv{\cal\bar{C}}$, ${\cal{C}}^{2}\equiv{\cal{C}}$,  ${\cal\bar{P}}_{2}\equiv{\cal\bar{P}}$  and ${\cal{P}}^{2}\equiv{\cal{P}}$. Second, taking the variations over  $p_{x}$, $p_{y}$, $p_{z}$, ${\cal{P}}$ and ${\cal\bar{P}}$ yields the following relations 
\begin{eqnarray}\label{28}
    p_{x}=\dot{x}+gyN,\qquad p_{y}=\dot{y}-gxN,\qquad  p_{z}=\dot{z}-N,\qquad {\cal\bar{P}}=- \dot{\bar{\cal{C}}},\qquad {\cal{P}}=\dot{\cal{C}}.
\end{eqnarray}
Finally, identifying $N$ with $\zeta$ (cf. \eqref{21}) and substituting Eqs. \eqref{28} in action \eqref{27}, we obtain the following BRST invariant effective action
\begin{eqnarray}\label{29}
  \Tilde{S}_{eff}&=&\int dt \,\Big[\frac{1}{2}\,\Big((\dot{x}+gy\zeta)^{2}+(\dot{y}-gx\zeta)^{2}+(\dot{z}-\zeta)^{2}\Big)+U(x^{2}+y^{2})\nonumber\\&+&B(\dot{\zeta}-z+\frac{1}{2}B)-{\dot{\bar{\cal{C}}}}\dot{\cal{C}}-\bar{\cal{C}}{\cal{C}}\Big].
\end{eqnarray}
Corresponding BRST symmetry transformations $(\Tilde{\delta}_{b})$ are listed below
\begin{eqnarray}\label{33}
 \Tilde{s}_{b} x=-gy{\cal{C}},\quad 
    \Tilde{s}_{b} y=gx{\cal{C}},\quad\Tilde{s}_{b} z={\cal{C}}, \quad
      \Tilde{s}_{b}\zeta=\dot{\cal{C}}, \quad \Tilde{s}_{b}{\cal{C}}=0,\quad\Tilde{s}_{b}\bar{\cal{C}}=-B,\quad\Tilde{s}_{b} B=0,
\end{eqnarray}
which leaves the action $(\tilde S_{eff})$ invariant.

It is worthwhile to mention that, if we perform path integration over $\cal{C}$ and $\cal\bar{C}$ instead of ${\cal{P}}$ and ${\bar{\cal{P}}}$ in Eq. \eqref{27}, the resulting effective action turns out to be
\begin{eqnarray}
  \Tilde{S}^{\prime}_{eff}&=&\int dt \,\Big[\,\frac{1}{2}\,\Big((\dot{x}+gy\zeta)^{2}+(\dot{y}-gx\zeta)^{2}+(\dot{z}-\zeta)^{2}\Big)+U(x^{2}+y^{2})\nonumber\\&+&B(\dot{\zeta}-z+\frac{1}{2}B)-{\dot{\bar{\cal{P}}}}\dot{\cal{P}}-\bar{\cal{P}}{\cal{P}}\,\Big].
\end{eqnarray}
This form of effective action can also obtained by the  transformation $\cal{C}\longrightarrow\cal{P}$ and
$\cal\bar{C}\longrightarrow\cal\bar{P}$ in Eq. \eqref{29} and vice-versa.

\section{Physicality criteria: BRST charge}\label{s4}
We note that, in both the polar and Cartesian coordinates, the conserved BRST charges
\begin{eqnarray}\label{charge}
    Q={\cal{C}}\,(gp_{\theta}+p_{z})\,+\,\dot{\cal{C}}p_{\zeta}, \qquad \Tilde{Q}={\cal{C}}\big(g\,(xp_{y}-yp_{x})+p_{z}\big)+\dot{\cal{C}}p_{\zeta},
\end{eqnarray}
are the generators of BRST symmetry transformations (cf. \eqref{BRST1} and \eqref{33}) under which the effective actions $(S_{eff}$ and $ \tilde S_{eff})$ remain quasi-invariant as
\begin{eqnarray}
    s_{b} S_{eff}=\frac{d}{dt}(B{\dot{\cal{C}}}),\qquad \Tilde{s}_{b} \Tilde{S}_{eff}=\frac{d}{dt}(B{\dot{\cal{C}}}),
\end{eqnarray}
respectively. It is worth mentioning that, in Eq. \eqref{charge}, we have used the Euler-Lagrange equations of motion ${\cal{P}}=\dot{\cal{C}}$ and the expression for momenta $p_{\zeta}=B$. Now, it is straightforward to prove the conservation of BRST charge $Q$ and $\Tilde{Q}$ with the help of Euler-Lagrange equations of motion.  
These charges are also nilpotent of order two (cf. Eq. \eqref{11})
which signifies its fermionic character, since only operators with odd Grassmann parity can anti-commute with themselves to zero.%These charges are also nilpotent of order two (cf. \eqref{11}). 

In the BRST-quantized theory the total Hilbert space contains both the physical and ghost states as the direct product. Moreover, the ghost variables do not have any interactions with the physical variables of the system as they are decoupled. So, as the requirement of the physicality criteria, the dynamically stable subspace of states is annihilated by the above mentioned charges, i.e.;
\begin{eqnarray}
      Q\ket{\psi}=0, \qquad   \tilde Q |{\tilde \psi} \rangle=0,
\end{eqnarray}
which leads to the following conditions, respectively
\begin{eqnarray}
(gp_{\theta}+p_{z})\ket{\psi}=0, \qquad p_{\zeta}\ket{\psi}=0, \\
\big(g\,(xp_{y}-yp_{x})+p_{z}\big)\,|\tilde{\psi}\rangle=0, \qquad p_{\zeta}\,|\tilde{\psi}\rangle=0.
\end{eqnarray}
It is important to note that $(gp_{\theta}+p_{z})\approx0$ and $p_{\zeta}\approx0$ are first-class constraints of the system in the polar coordinates, whereas $\big(g\,(xp_{y}-yp_{x})+p_{z}\big)\, \approx 0$ and $p_{\zeta}\,\approx 0$ are first-class constraints in Cartesian coordinates. Therefore, physical states are annihilated by first-class constraints, which is consistent with the results in \cite{FLPR3} and the Dirac formalism.

\section{Generalized BRST symmetry of FLPR model}
We start with the partition function for the BFV-BRST invariant effective action of the FLPR model as
\begin{eqnarray}
     {\cal{Z}}=\int {\cal{D}}r\,{\cal{D}}\theta\,{\cal{D}}z\,{\cal{D}}\zeta\,{\cal{D}}B\,{\cal{D}}\bar{\cal{C}}\,{\cal{D}}{\cal{C}}\,e^{i\,S_{eff}},
\end{eqnarray}
where the effective action $(S_{eff})$ is given by
\begin{eqnarray}\label{F2}
    S_{eff}&=&\int dt\, \Big[\frac{1}{2}\,\dot{r}^{2}+\frac{1}{2}\,r^{2}\big(\dot{\theta}-g\zeta \big)^{2}+\frac{1}{2}\big(\dot{z}-\zeta\big)^{2}-V(r)\nonumber\\&+& B\Big(\dot{\zeta}-z+\frac{1}{2}\,B\Big)-{\dot{\bar{\cal{C}}}}\,{\dot{\cal{C}}}-{\cal{\bar{C}}}\,{\cal{C}}\Big].
\end{eqnarray}
This partition function and the effective action are invariant under the following off-shell nilpotent infinitesimal BRST transformations $(\delta_{b})$
\begin{eqnarray}
    {\delta}_{b} r=0,\quad
    {\delta}_{b} \theta=g{\cal{C}}\Lambda,\quad{\delta}_{b} z={\cal{C}}\Lambda,
      \quad {\delta}_{b}\zeta=\dot{\cal{C}}\Lambda,\quad
   {\delta}_{b}{\cal{C}}=0,\quad {\delta}_{b}{\cal\bar{C}}=-B\Lambda,\quad{\delta}_{b} B=0,
\end{eqnarray}
where $\Lambda$ is an infinitesimal global parameter. As long as this parameter $\Lambda$ is anti-commuting and space-time independent, most of the characteristics of BRST transformations are unaffected by the field dependency or order of magnitude of the parameter $\Lambda$. However, we can generalize the BRST transformations to FFBRST transformations by making the anti-commuting infinitesimal parameter finite and field dependent. First, to make the parameter $\Lambda$ field dependent, we introduce a continuous arbitrary parameter $\kappa\,(0\leq\kappa\leq1)$ such that 
\begin{equation}
    \phi(t, \kappa = 0) = \phi(t),\qquad \phi(t, \kappa = 1) = \phi'(t),
\end{equation}
where $\phi(t)$ denotes the generic variables of the theory. Now, the field-dependent infinitesimal BRST transformations can be written as
\begin{eqnarray}\label{inf}
    &&\frac{d\theta(t,\kappa)}{d\kappa}=g{\cal{C}}\Theta'[\phi(t, \kappa)],\quad
    \frac{dz(t,\kappa)}{d\kappa}={\cal{C}}\Theta'[\phi(t, \kappa)],\quad \frac{d\zeta(t,\kappa)}{d\kappa}=\dot{\cal{C}}\Theta'[\phi(t, \kappa)],\nonumber\\
    &&\frac{d\bar{\cal{C}}(t,\kappa)}{d\kappa}=-B\Theta'[\phi(t, \kappa)],\quad \frac{dr(t,\kappa)}{d\kappa}=0,\quad
    \frac{d{\cal{C}}(t,\kappa)}{d\kappa}=0,\quad  \frac{dB(t,\kappa)}{d\kappa}=0,
\end{eqnarray}
where $\Theta'[\phi(t, \kappa)]$ is a field dependent infinitesimal parameter. Then, we construct the FFBRST transformations $(\delta_{f})$ by integrating infinitesimal transformations in Eq. \eqref{inf} from $\kappa=0$ to $\kappa=1$ as follows:
\begin{eqnarray}\label{ff}
&& \delta_f\, \theta =g\,{\cal{C}}\,\Theta[\phi(t)], 
\qquad \delta_f\, z = {\cal C} \,\Theta[\phi(t)],\qquad \delta_f \zeta =\dot{\cal{C}}\,\Theta[\phi(t)],\nonumber\\
&&\delta_f \bar{\cal{C}} =-B\,\Theta[\phi(t)],\qquad \delta_f \,r =0, \qquad \delta_f B=0,\qquad \delta_f {\cal{C}}=0,
\end{eqnarray}
where the finite field-dependent parameter
\begin{eqnarray}
\Theta [\phi(t)] = \int^1_0 d\kappa' \Theta'[\phi(t, \kappa')]. 
\end{eqnarray}
The effective action $(S_{eff})$ remains invariant under the FFBRST transformations (i.e., $\delta_f\, S_{eff} = 0$). However, since the Grassmannian parameter is field-dependent, the path integral measure $\displaystyle {\prod_\phi} {\cal D}\phi(t)$ is not invariant under generalized BRST transformations. The Jacobian $J(\kappa)$ varies non-trivially and can be replaced by
\begin{eqnarray}
 J(\kappa) \mapsto   e^{-iS_1[\phi(t, \kappa), \kappa]},
\end{eqnarray}
if and only if the condition
\begin{eqnarray}\label{cond}
\int \prod_\phi {\cal D}\phi(t, \kappa) \bigg[\dfrac{1}{J(\kappa)}\dfrac{dJ(\kappa)}{d \kappa} - i\, \dfrac{d}{d\kappa} S_1[\phi(t, \kappa), \kappa] \bigg] \, e^{i(S_{eff} + S_1)} = 0,  
\end{eqnarray}
is satisfied \cite{FF1}. Here, $S_1[\phi(t, \kappa), \kappa]$ is some local functional of fields with boundary condition $S_1[\phi(t, \kappa), \kappa]_{\kappa = 0} = 0$ and 
\begin{eqnarray}\label{F10}
\dfrac{1}{J(\kappa)}\dfrac{dJ(\kappa)}{d \kappa} = - \int dt\, \sum_\phi \bigg[\pm\, s_b \,\phi(t, \kappa)\dfrac{\Theta'[\phi(t, \kappa)]}{\delta \phi(t, \kappa)}\bigg],
\end{eqnarray}
where $\pm$ is the sign for bosonic and fermionic fields, respectively. Now, we construct the field-dependent BRST parameter as
\begin{eqnarray}\label{F11}
\Theta'[\phi(t, \kappa)] = i \int dt\, \big(\dot{\zeta} (t, \kappa) - z(t, \kappa) \big)\, \bar {\cal C}(t, \kappa).     
\end{eqnarray}
Using Eq. \eqref{ff} and \eqref{F11}, the expression \eqref{F10} turns into
\begin{eqnarray}\label{F12}
\dfrac{1}{J(\kappa)}\dfrac{dJ(\kappa)}{d \kappa} &=& - i \int dt \Big[ B(t, \kappa) \big(\dot \zeta(t, \kappa) - z(t, \kappa)\big) + \bar {\cal C}(t, \kappa) \, {\cal C}(t, \kappa)\nonumber\\& +& {\dot {\bar {\cal C}}}(t, \kappa)\, \dot {\cal C}(t, \kappa)\Big].    
\end{eqnarray}
Further, we make an ansatz for $S_1[\phi(t, \kappa), \kappa]$ as follows:
\begin{eqnarray}\label{F13}
S_1[\phi(t, \kappa), \kappa] &=& i \int dt \Big[\xi_1(\kappa)\, B(t, \kappa)\, \dot \zeta(t, \kappa) + \xi_2(\kappa)\, B(t, \kappa)\, z(t, \kappa) + \xi_3\,(\kappa) \bar {\cal C}(t, \kappa) \, {\cal C}(t, \kappa) \nonumber\\
&+& \xi_4(\kappa)\, {\dot {\bar {\cal C}}}(t, \kappa)\, \dot {\cal C}(t, \kappa)   \Big],    \end{eqnarray}
where the coefficient $\xi_{i}\,(i=1,2,3,4)$ are arbitrary $\kappa$-dependent constant parameters. Now, differentiating $S_1[\phi(t, \kappa), \kappa]$ with respect to the parameter $\kappa$ by employing Eq. \eqref{inf}, we obtain
\begin{eqnarray}\label{F14}
\frac{dS_{1}}{d\kappa}&=&i\int dt\Big[\frac{d\xi_{1}(\kappa)}{d\kappa}\,B(t,\kappa)\,\dot{\zeta}(t,\kappa)+\frac{d\xi_{2}(\kappa)}{d\kappa}\,B(t,\kappa)\,z(t,\kappa)+\frac{d\xi_{3}(\kappa)}{d\kappa}\,{\cal\bar{C}}(t,\kappa)\,{\cal{C}}(t,\kappa)\nonumber\\
&+&\frac{d\xi_{4}(\kappa)}{d\kappa}\,{\dot{\bar{\cal{C}}}}(t,\kappa)\,\dot{\cal{C}}(t,\kappa)+\xi_{1}(\kappa)\,B(t,\kappa)\,\ddot{{\cal{C}}}(t,\kappa)\,\Theta'+\xi_{2}(\kappa)\,B(t,\kappa)\,{\cal{C}}(t,\kappa)\,\Theta'\nonumber\\
&+&\xi_{3}(\kappa)\,B(t,\kappa)\,{\cal{C}}(t,\kappa)\,\Theta'+\xi_{4}(\kappa)\,\dot{B}(t,\kappa)\,\dot{\cal{C}}(t,\kappa)\,\Theta']\Big].
\end{eqnarray}
Substituting Eqs. \eqref{F12} and \eqref{F14} into the necessary condition Eq. \eqref{cond} leads to
\begin{eqnarray}
\int dt \Big[\Big(\frac{d\xi_{1}}{d\kappa}-i\Big)B\,\dot{\zeta}+\Big(\frac{d\xi_{2}}{d\kappa}+i\Big)B\,z+\Big(\frac{d\xi_{3}}{d\kappa}-i\Big){\cal\bar{C}}\,{\cal{C}}+\Big(\frac{d\xi_{4}}{d\kappa}-i\Big){\dot{\bar{\cal{C}}}}\,\dot{\cal{C}}\nonumber\\
+\,\big(\xi_{2}+\xi_{3}\big)B\,{\cal{C}}\,\Theta'
+\big(\xi_{4} - \xi_{1}\big) \dot{B}\, \dot{\cal{C}}\,\Theta'\Big]=0.
\end{eqnarray}
Solving the above expression leads to a set of first-order linear differential equations as
\begin{eqnarray}
   \frac{d\xi_{1}}{d\kappa}-i=0,\qquad \frac{d\xi_{2}}{d\kappa}+i=0,\nonumber\\
   \frac{d\xi_{3}}{d\kappa}-i=0,\qquad \frac{d\xi_{4}}{d\kappa}-i=0,
\end{eqnarray}
and $\Theta'$-dependent non-local terms vanish for
\begin{eqnarray}
 \big(\xi_{2}+\xi_{3}\big)=\big(\xi_{4}-\xi_{1}\big)=0. 
\end{eqnarray}
The solutions of the differential equations under the initial condition $\xi_{i}(\kappa=0)=0$ is obtained as
\begin{eqnarray}
    \xi_{1}=i\,\kappa,\qquad \xi_{2}=-i\,\kappa,\qquad \xi_{3}=i\,\kappa,\qquad \xi_{4}=i\,\kappa.
\end{eqnarray}
Now, with these coefficients $\xi_{i}$, we can identify the functional $S_1[\phi(t, \kappa), \kappa]$ as follows:
\begin{equation}
S_1[\phi(t, \kappa), \kappa]=\int dt\,\kappa\, \big(- B\,\dot{\zeta}+ B\, z- \,\bar{\cal{C}}\,{\cal{C}}- \, {\dot{\bar{\cal{C}}}}\,\dot{\cal{C}}\big).
\end{equation}
Therefore, the total effective action, by adding this $S_1[\phi(t, \kappa), \kappa]$ to $S_{eff}$ in Eq. \eqref{F2}, can be written as follows
\begin{eqnarray}\label{FT}
    S_{eff}+S_{1}[\phi(t, \kappa), \kappa]&=&\int dt\, \Big[\,\frac{1}{2}\,\dot{r}^{2}+\frac{1}{2}\,r^{2}\big(\dot{\theta}-g\zeta \big)^{2}+\frac{1}{2}\big(\dot{z}-\zeta\big)^{2}-V(r)\nonumber\\&+&B\big(\dot{\zeta}-z+\frac{1}{2}B\big)-{\dot{\bar{\cal{C}}}}{\dot{\cal{C}}}-{\cal{\bar{C}}}{\cal{C}}+\kappa\big(-B\dot{\zeta}+Bz-\bar{\cal{C}}{\cal{C}}- {\dot{\bar{\cal{C}}}}\dot{\cal{C}}\Big].
\end{eqnarray}
Interestingly, at $\kappa=0$, the above action simplified to
\begin{eqnarray}
    S_{eff}+S_{1}[\phi(t,0),0]&=&\int dt\, \Big[\frac{1}{2}\,\dot{r}^{2}+\frac{1}{2}\,r^{2}\big(\dot{\theta}-g\zeta \big)^{2}+\frac{1}{2}\big(\dot{z}-\zeta\big)^{2}-V(r)\nonumber\\&+&B\Big(\dot{\zeta}-z+\frac{1}{2}B\Big)-{\dot{\bar{\cal{C}}}}{\dot{\cal{C}}}-{\cal{\bar{C}}}{\cal{C}}\Big],
\end{eqnarray}
which is same as the BFV-BRST gauge-fixed effective action in Eq. \eqref{F2}. However, at $\kappa=1$ the effective action in Eq. \eqref{FT} reduces to the gauge-invariant classical action of the model as given below 
\begin{eqnarray}
S_{eff}+S_{1}[\phi(t,1),1]&=&\int dt\, \Big[\frac{1}{2}\,\dot{r}^{2}+\frac{1}{2}\,r^{2}\big(\dot{\theta}-g\zeta \big)^{2}+\frac{1}{2}\big(\dot{z}-\zeta\big)^{2}-V(r)\Big].
\end{eqnarray}
In the above, we eliminate the Nakanishi-Lautrup auxiliary variable $B$ and (anti-)ghost variables $(\bar{\cal{C}}){\cal{C}}$ using Gaussian integration. Thus, we demonstrate that, with a judicious choice of the parameter, FFBRST transformations connect the BFV-BRST gauge fixed action and the classical gauge-invariant action of the FLPR model.

At this juncture, it is worth mentioning that the finite field-dependent parameter in Cartesian coordinates is the same as that for polar coordinates. Therefore, the computation in the Cartesian coordinates can be performed in a similar fashion.

\section{Conclusions}\label{sec4}
We have performed the BFV quantization of the FLPR model in both polar and Cartesian coordinates using admissible gauge conditions. The FLPR model is endowed with first-class constraints \textit{\`{a} la} Dirac's prescription for classification of constraints. In order to apply the BFV formalism, we have introduced a pair of ghost and anti-ghost variables along with auxiliary variables. The conserved and nilpotent  charge has been constructed using constraints. Whereas the fermionic gauge-fixing function is being put together by means of admissible gauge-fixing conditions. We have obtained the effective action, in the extended phase space, by the usual Legendre transformation. By eliminating a few variables using Gaussian integration, we have established the BRST invariant gauge-fixed effective action and corresponding BRST symmetries of the FLPR model in both polar and Cartesian coordinates. We have also indicated that the effective action has a discrete symmetry in the ghost sector. Using the physicality criterion, we have shown that the physical states of the theory are annihilated by the first-class constraints, which is consistent with the Dirac formalism.

Further, we have generalized the infinitesimal BRST symmetries by making the transformation parameter finite and field dependent, thereby establishing the FFBRST symmetries of the FLPR model. Although the action is invariant under such transformations, the field dependence of the finite parameter breaks the invariance of generating functional. We have chosen an appropriate FFBRST parameter and computed the non-trivial Jacobian of the FFBRST transformation. Thus, we are able to show that with a judicious choice of the finite field-dependent parameter, FFBRST transformations can bridge between generating functional corresponding to two different actions. We have demonstrated the interconnection between the BFV-BRST gauge-fixed effective action and the classical gauge invariant action within FFBRST formulation.  \\

\noindent
{\bf Acknowledgment:} SG would like to dedicate this paper in the fond memories of Prof. Victor O. Rivelles who was a kind mentor.

\end{document}